\documentclass[prl,aps,twocolumn,showpacs]{revtex4}

\usepackage{graphicx}
\usepackage{dcolumn}
\usepackage{bm}

\begin{document}


\title{'Hole-digging' in ensembles of tunneling Molecular Magnets}

\author{I. S. Tupitsyn$^{1,2}$, P. C. E. Stamp$^{1}$ and N.V. Prokof'ev$^{2,3}$ }
\affiliation{ $^{1}$ Physics Department, and Canadian Institute for Advanced Research, \\
University of British Columbia, 6224 Agricultural Rd., Vancouver BC, Canada V6T 1Z1. \\
$^{2}$ Russian Research Center "Kurchatov Institute", Kurchatov
Sq.1, Moscow 123182, Russia. \\
$^{3}$ Department of Physics, University of Massachussets,
Amherst, MA 01003, USA \\ }


\begin{abstract}

The nuclear spin-mediated quantum relaxation of ensembles of
tunneling magnetic molecules causes a 'hole' to appear in the
distribution of internal fields in the system. The form of this
hole, and its time evolution, are studied using Monte Carlo
simulations. It is shown that the line-shape of the tunneling hole
in a weakly polarised sample must have a Lorentzian lineshape- the
short-time half-width $\xi_o$ in all experiments done so far
should be $\sim E_0$, the half-width of the nuclear spin
multiplet. After a time $\tau_o$, the single molecule tunneling
relaxation time, the hole width begins to increase rapidly. In
initially polarised samples the disintegration of resonant
tunneling surfaces is found to be very fast.

\end{abstract}

\maketitle

\vspace{1cm}


In the last few years many experiments have been done on the
tunneling relaxation of magnetic molecules
\cite{can00,BB00,WWrev,kent02,boka00,chio00}. These couple to each
other, via dipole and weak exchange interactions, and also to
phonon \cite{pol96} and nuclear \cite{PS96} environments. The most
complete results have appeared on the $Fe$-8 system, where
measurements of the relaxation \cite{can00,WWrev,OSP,WWPRL,WWISO}
were interpreted using a theory in which the tunneling dynamics
was controlled by dynamic hyperfine and dipolar fields
\cite{PSPRL}. A key connection between theory and experiment is
the existence of 'hole-digging' in the distribution $M(\xi,t)$
of magnetisation $M$ over longitudinal bias energy $\xi$ in the low
$T$ regime where the molecules behave as 2-level systems and phonons
are unimportant \cite{WWrev,WWPRL,PSPRL,PSSB}. The hole appears
because a molecule can then only tunnel when in resonance. Without
nuclear spins resonance only occurs if the local energy bias
$\vert \xi \vert \lesssim \Delta_o$. However the nuclear spins
mediate inelastic tunneling over a much larger range $\vert \xi
\vert < \xi_o$; in $Fe$-8 at low fields, $\xi_o/\Delta_o \sim
10^5$. Theory thus predicted an initial holewidth $\xi_o$,
provided $\xi_o \ll$ the range of energy bias caused by the
dipolar field distribution. The experimental half-width $w_o$ of
the hole at {\it short times}, in initially strongly annealed
samples, was interpreted \cite{WWrev,WWPRL} to be the total
half-width $E_o$ of the multiplet of nuclear spin states coupled
to the molecule- this latter can be calculated if the hyperfine
couplings are known \cite{PS96}. Support for this interpretation
came when isotopes were varied \cite{WWISO}; $w_o$ agreed both
with approximate calculations of $E_o$, for different sets of
nuclear isotopes, and with values of $\xi_o$ extracted from
independent measurements of the sample relaxation rate
$\tau_Q^{-1}$.

It is important to understand this hole-digging, not only to
evaluate the claim that it demonstrates the role of nuclear spins
in controlling the molecular dynamics (and the use of the holes as
diagnostic tools), but also because the physics of the hole
formation turns out to be rather interesting. There are 4 obvious
questions which one may ask:

(i) Why is $\xi_o = E_o = w_o$ in the $Fe$-8 experiments, when
$M(0)$ is small?

(ii) The experimental hole width $w(t)$ varies in time (with $w(t)
\rightarrow w_o$ at short times) - how and why?

(iii) $w(t)$ also depends very strongly on the initial
polarisation $M(t=0)$ in the system ($M(t)=\int d\xi M(\xi,t)$), but
$w_o$ is apparently independent of $M(0)$ for small $M(0)/M_s$; is
this expected theoretically?

(iv) Why is the experimental lineshape Lorentzian?

In fact most published experiments are on partially annealed
samples, where $M(0)/M_s < 1$ ($M_s$ being the saturated
magnetisation of a fully polarised sample- henceforth we measure
$M(t)$ in units of $M_s$). In the theory \cite{PS96,PSPRL,PSSB}
the quantity $\xi_o$ was not necessarily the same as $E_o$; and
neither the hole lineshape, nor the dependence of $w(t)$ on $M(0)$
or $t$, were given. Moreover the theory only deals with weakly or
strongly polarised samples, whereas almost all experiments are on
partially annealed samples, with intermediate polarisation- these
are beyond the reach of analytic work.

To address these 4 questions, we present here numerical results
for the hole lineshape, and the holewidth $w(t)$ as a function of
t and initial annealing. We also calculate results for the $Fe$-8
system, to provide predictions which can be compared with
experiment.

\vspace{3mm}

{\it 1. Short-time hole width}: The experimental interpretation,
that at short times $w_o = E_o$, appears to be based on the
assumption (made explicitly in ref. \cite{alonso}) that the
nuclear bias potential behaves as a noise, fluctuating rapidly
over the whole nuclear multiplet (halfwidth $E_o$). Actually this
assumption is unnecessary- in many cases \cite{holeW} the hole
width depends on the energy exchanged between nuclear spins and
the molecules during inelastic tunneling \cite{PSSB}. If the
strength of the hyperfine interactions between the molecular spin
${\bf S}$ and the $k$-th nuclear spin ${\bf I}_k$ is
$\omega_k^{\parallel}$, then $E_o^2 = \sum_k
(\omega_k^{\parallel})^2 (I_k+1)I_k/3$. However energy is
exchanged over a range $\xi_o \sim \kappa \omega_o$, where
$\omega_o = N^{-1} \sum_k \omega^{||}_k$ is the mean hyperfine
coupling to the $N$ nuclear spins in the molecule. In low fields
$2\kappa = \sum (\omega_k^{\perp}/\omega_k^{\parallel})^2$, where
$\omega_k^{\perp} = g_k^N \mu_N H_{o}$ is the Zeeman coupling to a
field $H_{o}$. In experiments with zero applied field, $H_{o}$ is
just the strength of the local dipolar field from the other
molecules- its root mean square in nearly annealed $Fe$-8
molecules is $\sim 0.05~T$. If $\kappa \gtrsim \sqrt{N}$, then
\cite{PSSB} $\xi_o \rightarrow E_o$.

\begin{figure}[h]
\centering
\vspace{-2.8cm}
\hspace{0.cm}
\includegraphics[scale=0.4]{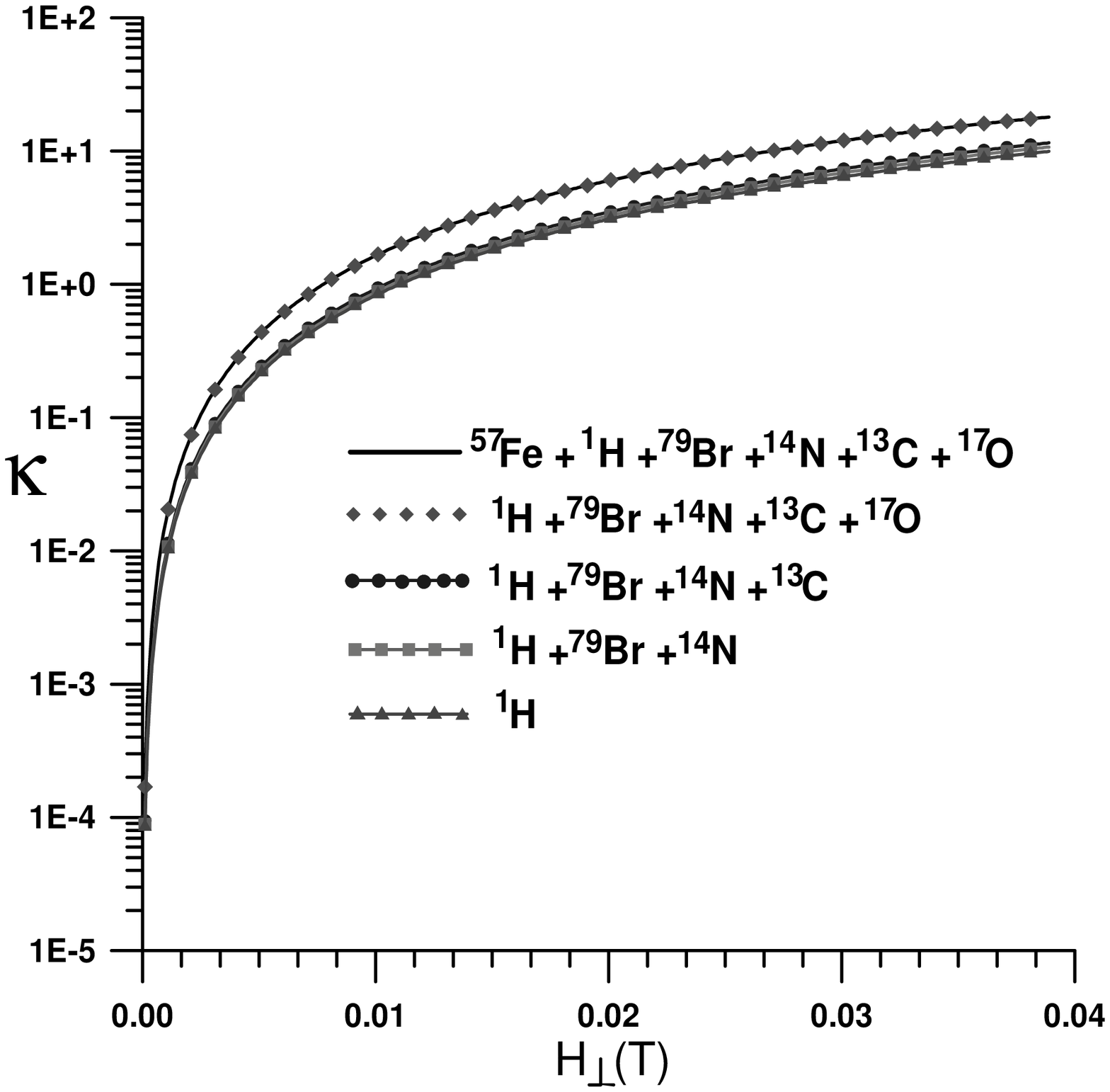}
\vspace{-2.2cm}
\caption{The variation of $\kappa({\bf
H}_o^{\perp})$ with transverse field ${\bf H}_o^{\perp}$,
calculated for a single $Fe$-8 molecule at low fields, for various
isotopic mixtures. The hyperfine interactions were assumed to be
dipolar, except for the core polarisation coupling to $^{57}Fe$,
which was taken from experiment.}
\label{fig:fig1}
\end{figure}

\vspace{-0.3cm}

Evaluation of $\kappa$ at low fields is simple for $Fe$-8 if we
assume that all hyperfine couplings are dipolar except those to
$^{57}Fe$ nuclei \cite{WWISO}, since all nuclear positions and
moments are known \cite{CDC}. The couplings $\{
\omega_k^{\parallel} \}$, $\{ \omega_k^{\perp} \}$ can then be
computed \cite{rose}. Fig. 1 shows the results, for different
isotopic concentrations in $Fe$-8. We see that if it were not for
the strong intermolecular dipolar fields, $\xi_o$ would be much
less than $E_o$; however, in fields $\sim 0.05~T$, they will be
very close to each other.

\vspace{3mm}

{\it 2. Time dependence of holewidth}: To find the time dependence
of $w(t)$ we perform Monte Carlo simulations for the time
evolution of $M(\xi,t)$ on a simple cubic lattice, averaging over
several hundred random initial spin configurations consistent with
the initial $M(t=0)$ (for previous MC simulations see refs.
\cite{PSPRL,VillSurf} (polarised systems) and \cite{alonso}
(strongly annealed systems)). Individual spins, modelled as point
dipoles oriented along $\pm \hat{z}$, relax incoherently in a
local bias field $\xi$ at a rate $\tau^{-1}_N(\xi) = \tau^{-1}_0
e^{- |\xi| / \xi_0}$, where $\tau_o^{-1} \propto
\Delta_o^2/\Lambda_N$, with $\Delta_o$ the tunneling matrix
element and $\Lambda_N$ a nuclear spin diffusion rate
\cite{PSPRL}. The dipolar fields at each site, calculated
numerically, are updated at time intervals $\delta t \ll \tau_o$
(typically $\delta t/\tau_o \sim 10^{-3}$ at the beginning of the
decay) by flipping individual spins with probability $1-e^{-
\delta t / \tau_N (\xi)}$. We define in the usual way the
magnetisation distribution $M(\xi, t) = \int d^3r
[P_{\uparrow}(\xi, {\bf r}; t) - P_{\downarrow}(\xi, {\bf r}; t)]$
for the sample, where $P_{\sigma}(\xi, {\bf r}; t)$ is the
probability that a molecule at position ${\bf r}$ with spin
orientation $\sigma = \uparrow,\downarrow$ will be in bias field
$\xi$ at time $t$. A typical result is shown in Fig. 2.

\begin{figure}[h]
\centering
\vspace{-2.4cm}
\hspace{0.0cm}
\includegraphics[scale=0.4]{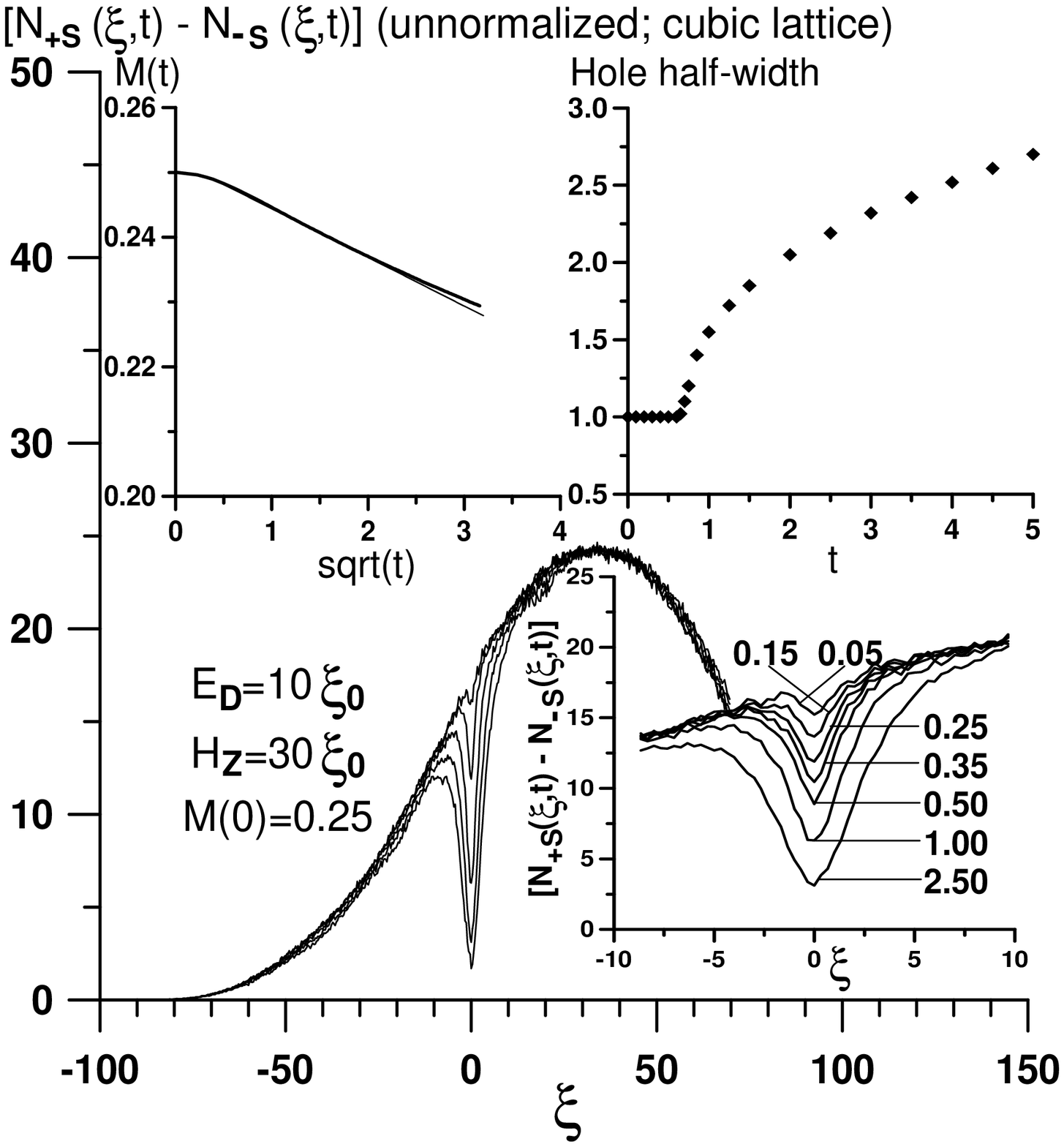}
\vspace{-2.4cm}
\caption{The tunneling hole dynamics, obtained by
Monte Carlo simulations, performed on a model cubic lattice of
$40^3$ point dipoles ($E_D$ is the strength of the nearest
neighbour dipolar interaction, and $E_D/\xi_o$ here). The main
graph shows the time evolution of $M(\xi,t)$, defined in the text,
for a sample in external field $H_z = 30 \xi_o$ at $M(0)=0.25$
(for which $M(\xi,0)$ is nearly Gaussian, with a halfwidth $W_D \sim
4.4E_D$ for this cubic lattice). The two upper insets are (i) at
left, the magnetization curve $M(t)$ vs the $\sqrt{t}$ and (ii) at
right, the tunneling hole half-width $w(t)$ as a function of time.
The inset at the bottom right enlarges the region around $\xi=0$
to show the hole dynamics. Note that time $t$ is shown in units of
$\tau_o$ and bias $\xi$ in units of $\xi_o$.}
\label{fig:fig2}
\end{figure}

A number of important general features of these results are as
follows. If {\it and only if} the Gaussian halfwidth $W_D$ of
$M(\xi,t=0)$ satisfies $W_D \gg \xi_o$, we find:

(i) for $t < \tau_0$ the hole halfwidth $w(t) \rightarrow w_o$,
and $w_o = \xi_o$ identically. Given the result above, that $\xi_o
\sim E_o$, this means the short-time holewidth $w_o$ is indeed
measuring $E_o$, as previously assumed \cite{WWPRL,WWISO,WWEPL}.

(ii) When $t \gtrsim \tau_0$ the hole rapidly broadens. The sharp
rise in $w(t)$ when $t \sim \tau_o$ appears not to be an artifact
of the simulation. We have not yet been able to establish the
asymptotic form of $w(t)$ for $\tau/\tau_o \gg 1$, but for
$1 < t / \tau_o < 5 $, $w(t)$ grows slower than $t^{0.4}$.

(iii) For $t > \tau_o$ a square root relaxation (ie., $\vert M(t)
- M(0) \vert \propto (t/\tau_Q)^{1/2}$) is always found, for any
$M(0)$. This is obeyed until the fractional change in $\vert
M(t)-M(0)\vert \sim 0.1-0.15$.

\vspace{3mm}

{\it 3. Hole lineshape}: When $M(0) < 0.5$, we find the hole
lineshape is Lorentzian. If $M(0) > 0.5$, deviations begin to
appear; when $M(0) \sim 1$ the lineshape is not Lorentzian at all.

It is important to understand the origin of this Lorentzian
lineshape. In a strongly annealed sample the dipolar field
fluctuates very rapidly from one molecule to another, but on a
coarse-grained scale the sample should look homogeneous. We can
therefore define the average time $\tau_{de}(\xi)$ for a molecule
with initial bias $\xi$ to come to resonance. The total
magnetization is then given by the homogeneous ensemble average:
\begin{equation}
M(t) =  M_{eq} + (M(0)-M_{eq}) \int d\xi N(\xi) e^{- t /
\tau_{de}(\xi)},
\label{MDS}
\end{equation}
where $M_{eq} = M(t \to \infty)$ and $N(\xi) = \int d^3r
[P_{\uparrow}(\xi, {\bf r}; 0) + P_{\downarrow}(\xi, {\bf r}; 0)]$
is the total dipolar field distribution. The Lorentzian hole
lineshape results directly from a Lorentzian form for
$\tau_{de}^{-1}(\xi)$. We demonstrate this by calculating
$\tau_{de}(\xi)$ using the Monte Carlo procedure.

\begin{figure}[h]
\centering
\vspace{-1.9cm}
\hspace{0.0cm}
\includegraphics[scale=0.35]{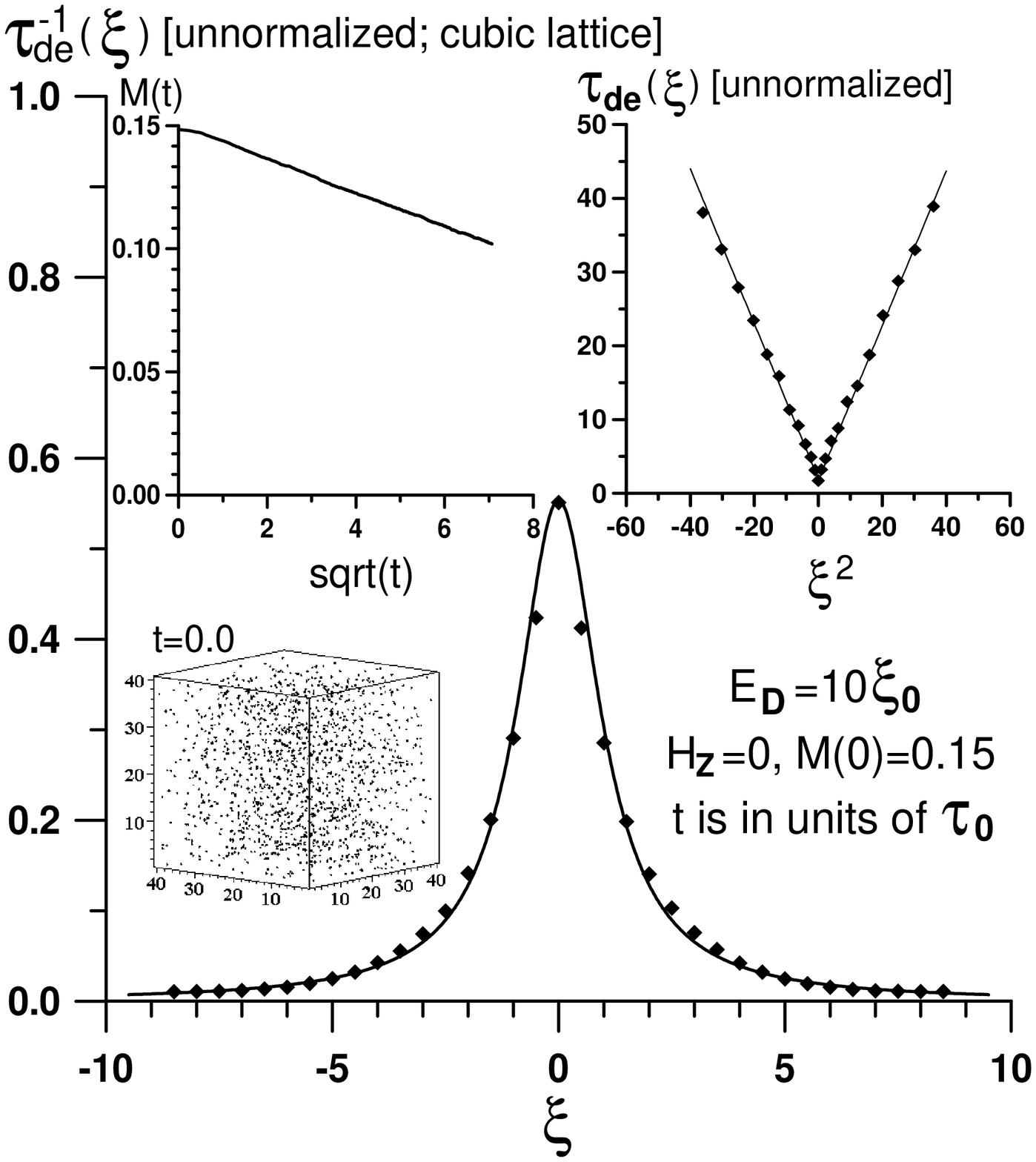}
\vspace{-1.9cm}
\caption{The average decay {\it rate}
$\tau^{-1}_{de}(\xi)$ obtained by the MC simulations in a cubic
sample of $40^3$ molecules (diamonds). By solid line we plot the
fitting Lorentzian curve with the half width $\xi_0$. Three insets
are: (i) the magnetization relaxation curve $M(t)$ vs the
$\sqrt{t}$; (ii) the average decay {\it time} $\tau_{de}(\xi)$ vs
the $\xi^2$ (diamonds); and (iii) the set of points in space,
where for $M(0)=0.15$ at $t=0$ molecules are in resonance.}
\label{fig:fig3}
\end{figure}

\vspace{-0.3cm}

At $t=0$ all molecules were split into different bias groups of
width $\xi_0/2$. When some molecule flipped, the number of
molecules in its group was reduced by one. The time taken for each
bias group to decrease its initial number of molecules by a factor
$e$ is assumed to be the decay time $\tau_{de}$ for this group.
Fig.3 shows the results for $M(0)=0.15$; one has a Lorentzian
line-shape with half-width $\xi_0$.

Note one simple implication of this result: when $t / \tau_0 > 1$,
the dominant contribution to the integral (\ref{MDS}) comes from
$\xi > \xi_0$ where $\tau_{de}(\xi) \sim \xi^2$; by making the
integral in (\ref{MDS}) dimensionless, we see the magnetisation
must decay like $\vert M(t)-M(0)\vert \sim \sqrt{t}$. Thus the
Lorentzian lineshape of the hole in $M(\xi)$, in strongly annealed
samples, is directly connected to the square root relaxation,
arising from the dipolar form of the intermolecular interactions.
Notice, incidentally, that the distribution of molecules in the
sample found to be in resonance for these strong annealings is
indeed almost homogeneous (Fig.3, inset bottom left).

\vspace{3mm}

{\it 4. Application to $Fe$-8 system}: Instead of point dipoles we
now have 8 spin-$5/2$ $Fe^{+3}$ ions correctly positioned and
oriented inside each molecular unit cell, arranged in a triclinic
lattice array. The lattice parameters, easy and hard axis
orientations, were taken from standard sources
\cite{OrientFe,CDC}. As an example we show in Fig. 4 results for
$M(t)$ and $\tau_{de}^{-1}(\xi)$, for an initial magnetisation
$M(0) = 0.2$.


\begin{figure}[h]
\centering
\vspace{-2.0cm}
\hspace{0.0cm}
\includegraphics[scale=0.35]{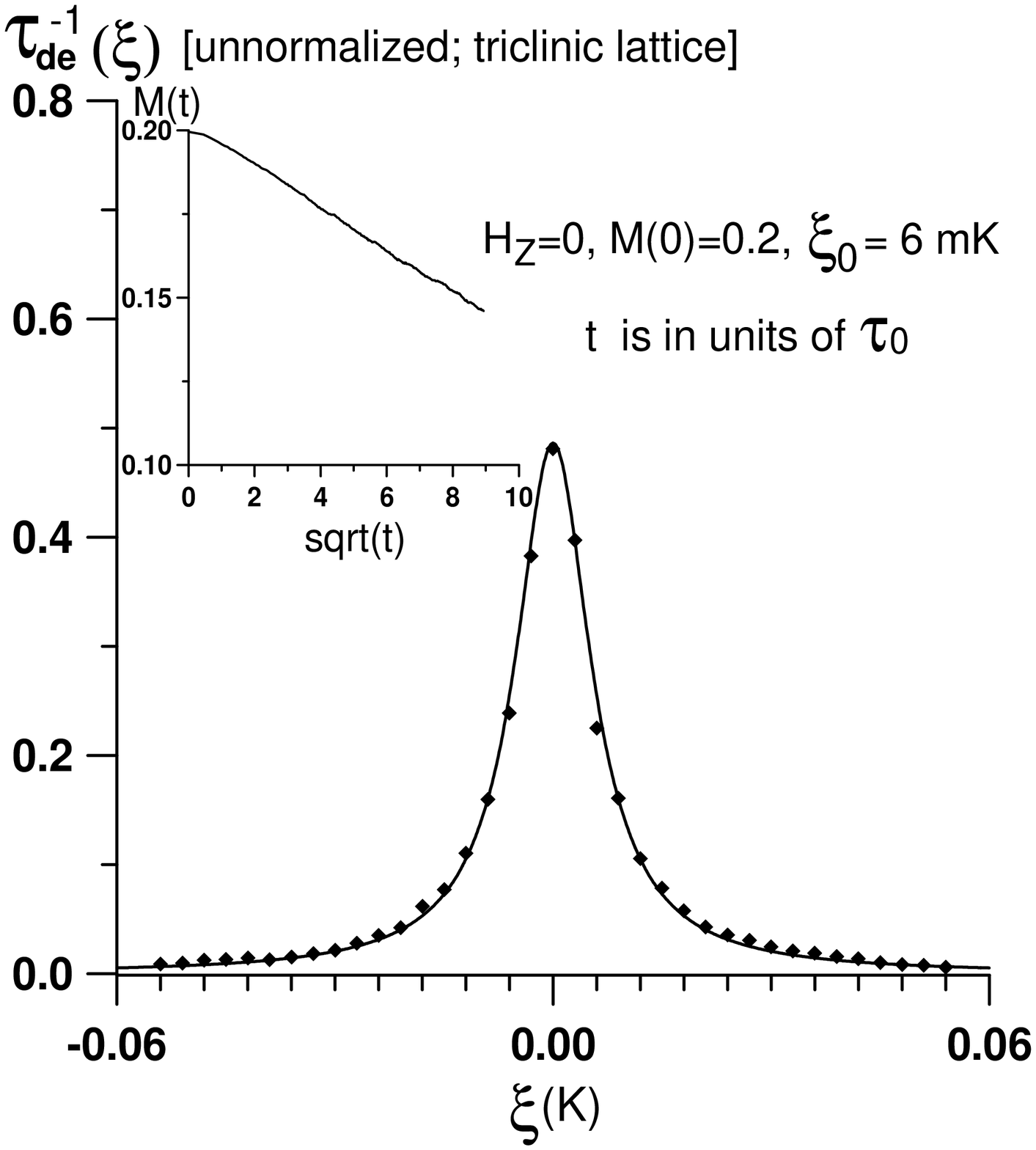}
\vspace{-1.8cm}
\caption{Plot of $\tau_{de}^{-1}$, as in Fig. 3,
but now for a crystal of $Fe$-8 molecules (with $40^3$ molecules
arranged in a triclinic lattice). The inset above shows the decay
of $M(t)$ with $\sqrt{t}$ for a partially annealed sample (initial
polarisation $M(0) = 0.2$). The lattice parameters for the $Fe$-8
triclinic lattice (a=10.522(7)\AA, b=14.05(1)\AA, c=15.(1)\AA,
with angles $\alpha=89.90(6), \beta=109.65(5), \gamma=109.27(6)$,
in degrees), and the positions of the 8 spin-$5/2$ ions in each
unit cell, were taken from ref. \cite{CDC}. The bias energy $\xi$
is now given in Kelvin units, and the value $\xi_o = 6~mK$ is that
appropriate to natural isotopic concentration \cite{WWISO}.}
\label{fig:fig4}
\end{figure}

\vspace{-0.3cm}

Our basic conclusion here is that none of the results found for
simple cubic lattices with point dipoles is essentially altered by
going to a magnetisation distributed over the 8 $Fe$ ions, or by
the change of lattice structure- one still has a Lorentzian
lineshape for $\tau_{de}^{-1}$, and the $\sqrt{t/\tau_Q}$
relaxation form for $M(t)$, for $M(0) \lesssim 0.5$, for the
simulations we have done \cite{sizeEff}.


\vspace{3mm}


{\it 5. Nearly polarised samples}: Finally we examine the opposite
limit of strong initial polarisation, which has already been
discussed both analytically and numerically \cite{PSPRL,VillSurf}.
In a sample of generic shape, the molecules that can tunnel
inelastically (ie., those satisfying the 'resonance' condition
$\vert \xi \vert < \xi_o$) form well-defined 'resonance surfaces'
in space if the sample is nearly polarised, and $\xi_o \ll E_D$
\cite{VillSurf,seattle98}. This is because now the dipolar field
is dominated not by short-range fluctuations in space, but by the
smoothly varying demagnetisation field. These resonance surfaces
can be visualised by plotting those molecules satisfying the
resonance condition at time $t$, during a Monte Carlo simulation.
The spatial variation of the longitudinal component of the dipolar
field can then be described by a slowly-varying demagnetisation
field $E({\bf r},t)$, plus a smaller rapidly varying fluctuation
$\delta \xi ({\bf r},t)$, having Lorentzian distribution
\cite{PSPRL,AndBer} over values of $\delta \xi$, and $\langle
\delta \xi ({\bf r},t) \rangle = 0$.

\begin{figure}[h]
\centering
\vspace{-1.9cm}
\hspace{0.5cm}
\includegraphics[scale=0.35]{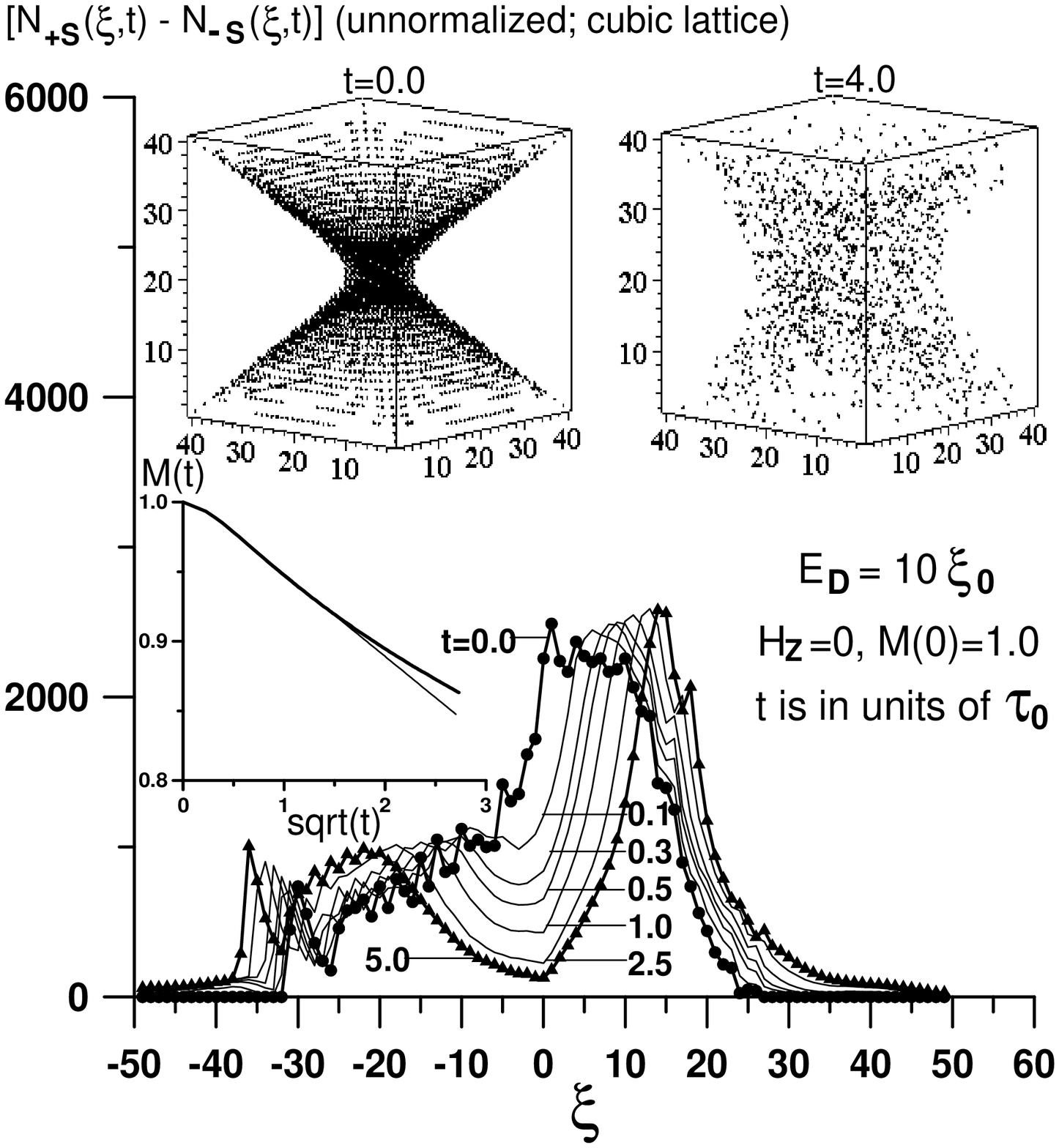}
\vspace{-1.9cm}
\caption{The main figure shows the time evolution
of the distribution $M(\xi,t)$ of magnetisation caused by dipolar
fields, in an initially polarized sample ($M(t=0)=1$); the same
cubic sample of Fig. 2 was used. The 3 insets show: (i) the
resonance surface at $t=0$, showing all molecules in the sample
satisfying the resonance condition in the text; (ii) the same
result at $t = 4 \tau_o$, where we see that the molecules
satisfying the resonance condition have now spread in space (iii)
the relaxation of the normalized magnetization $M(t)$ vs
$\sqrt{t}$ for the same sample (with the square root form shown as
a thin line).}
\label{fig:fig5}
\end{figure}

\vspace{-0.3cm}

It then becomes interesting to ask how these surfaces evolve in
time. Again we must resort to Monte Carlo simulations. Fig. 5
shows a typical example- for all sample geometries we have studied
the surfaces break up rapidly once the fraction $(1-M(t)) \sim
0.07-0.11$, corresponding closely with the onset of deviations
from the square root scaling. This is intriguing because it
indicates the deviations may be caused not only by higher
correlations between the molecules, as previously assumed
\cite{PSPRL}; this result needs further investigation. Notice that
the hole lineshape is now irrelevant to the square root form of
the relaxation dynamics, which comes instead from the Lorentzian
distribution of $\delta \xi ({\bf r},t)$ \cite{PSPRL,AndBer}. The
hole lineshape is now controlled by the sample shape.

Our essential conclusion in this paper is that one can use Monte
Carlo methods to investigate the relaxation characteristics in
regimes inaccessible to analytic theory. This allows one to
understand the lineshape, width and time evolution of the hole
that appears in $M(\xi,t)$.

We thank NSERC and the CIAR in Canada, and grant number
NS-1767.2003.2 in Russia, for support.



\begin{references}
\bibitem{holeW} The measured nuclear $T_1$ in systems like $Fe$-8
or $Mn$-12 is typically $\sim 10-100$ secs \cite{NMR}, longer than
the time it takes the hole to develop.
\bibitem{sizeEff} We found that size effects are non-negligeable for
samples smaller than $40^3$ (and they apparently depend strongly
on lattice symmetry).
\end{references}
\end{document}